\documentclass[]{aa}
\usepackage{graphicx}
\usepackage{txfonts}
\begin{document}
   \title{The Antares emission nebula and mass loss of $\alpha$ Sco A
   \thanks{Based on observations under program ID 076.D-0690(A) with the
   Ultraviolet and Visual Echelle Spectrograph (UVES) on the Very Large Telescope
   (VLT) Unit 2 (Kueyen) at Paranal, Chile, operated by ESO}}

   \author{D. Reimers
          \and H.-J. Hagen  \and R. Baade \and K. Braun
          }

   \institute{Hamburger Sternwarte, Universit\"at Hamburg,
         Gojenbergsweg 112, 21029 Hamburg, Germany
          }

   \offprints{D.~Reimers (dreimers@hs.uni-hamburg.de)}

   \date{received date; accepted date}


  \abstract
  {}
  {The Antares nebula is a peculiar emission nebula seen in numerous
[\ion{Fe}{ii}] lines and in radio free-free emission, probably
associated with the \ion{H}{ii} region caused by $\alpha$ Sco B in the
wind of $\alpha$ Sco A. High-resolution spectra with spatial
resolution were used to study the emission line spectrum, the physical
nature of the nebula and to determine the mass-loss rate of the M
supergiant $\alpha$ Sco A.}
  {The Antares nebula was mapped with long-slit (10$\arcsec$) and
high-resolution ($R=80\,000$) spectra using UVES at the VLT. The
resulting 2-D images were used  to
reconstruct a 3-D picture of the \ion{H}{ii} region and its absolute
location in space relative to $\alpha$ Sco A.}
 {We found that the Antares nebula shows, in
addition to numerous [\ion{Fe}{ii}] lines, the Balmer line
 recombination spectrum
H$_{\alpha}$, H$_{\beta}$ up to H$_{10}$, and [\ion{N}{ii}]
6583/6548~{\AA}, H$_{\alpha}$ and [\ion{N}{ii}]  with the same extent
 as seen in cm radio free-free
emission. Combining velocity information from optical and GHRS/HST
spectra with H$_{\alpha}$ velocities, the \ion{H}{ii} region is found
to be located $\sim 215$~AU behind the plane of the sky
of  $\alpha$ Sco A. From the
H$_{\alpha}$/[\ion{N}{ii}] intensity ratio and the non-visibility of
the [\ion{O}{ii}] 3726/3729~{\AA} lines we estimate a low mean electron
temperature of $\overline{T}_{\rm e} = 4900$~K and an N abundance
enhanced by a factor of $\sim$ 3 due to the CNO cycle in $\alpha$ Sco
A. The shape and size of the \ion{H}{ii} region yield a mean mass-loss
rate of (1.05 $\pm$ 0.3) $\times$ 10$^{-6}$ M$_{\odot}$\,yr$^{-1}$.
The [\ion{Fe}{ii}] lines originate predominantly
 at the edges (rear and front) of the \ion{H}{ii} region. 
UV continuum pumping as well as collisional excitation seem to be responsible
for the observed iron lines.
}
  {}


\maketitle
%

\section{Introduction}
The Antares nebula is a unique object in the sky. Antares
\mbox{($\alpha$ Sco A, M 1.5 Iab)} and its blue visual companion
$\alpha$ Sco B (B 2.5 V) are known to be surrounded by a circumstellar
envelope seen in absorption in both components which can be used to
determine the mass-loss rate of the M supergiant (Deutsch \cite{deu60};
 Kudritzki \& Reimers \cite{kud78}; Hagen et al. \cite{hag87};
Baade \& Reimers \cite{baa07}).
 The common envelope has also been known for a long
time as an emission nebula with strong [\ion{Fe}{ii}] lines (Wilson \& Sanford
\cite{wil37}; Struve \& Swings \cite{str40}; Swings \& Preston \cite{swi78}).
 The most extensive study was that by
Swings \& Preston (\cite{swi78}) based on high-resolution
long-slit photographic Coud\'e spectra taken with the Mount Wilson 100
inch and Palomar 200 inch telescopes. They found that the
``[\ion{Fe}{ii}]-rich nebula'' is strong roughly $3\farcs5$ around the B~star
surrounded by a zone of weaker lines which may extend in the NW-SE
direction up to $15\arcsec$. It has been shown by Kudritzki \& Reimers
(\cite{kud78}) that the [\ion{Fe}{ii}] emission lines are probably
associated with the \ion{H}{ii} region formed by the B star within the stellar
wind of the M supergiant.

The \ion{H}{ii} region has  been detected and resolved
 at cm radio wavelengths with the VLA
(Hjellming \& Newell \cite{hje83}). The maximum \ion{H}{ii} region radio
emission was found to be centered roughly $0\farcs5$ from the B star on the
line connecting the two stars.

Several questions remained unanswered. Why are the Balmer lines,
expected in emission due to recombination within the \ion{H}{ii} region,
absent or weak compared to the [\ion{Fe}{ii}] lines? Why are the ``classical''
\ion{H}{ii} region emission lines absent or weak? Why are the [\ion{Fe}{ii}] lines
double west of B, but single east of B (between the two stars)? With
the advent of UVES at the VLT it was obvious that progress in our
understanding of the kinematics of the Antares nebula is now possible
due to the high spectral resolution, high pointing accuracy, and
stability of UVES. We used the spectrograph at a resolution of 80\,000 to
map the nebula in 2-D with 23 slit positions covering the whole nebula
(Fig.~\ref{slitpos}).

In section 2 we describe the observations and the data reduction which
was difficult due to (unexpected) scattered light from
the M supergiant even 10$\arcsec$ away from the bright star. Section 3
gives a phenomenological description of the observations. In section
4 we present the analysis of the UVES spectra as well as a discussion
of the obtained results.

\section{Observations and data reduction}
\subsection{Observations}
\begin{figure}
  \resizebox{\hsize}{!}{\includegraphics*{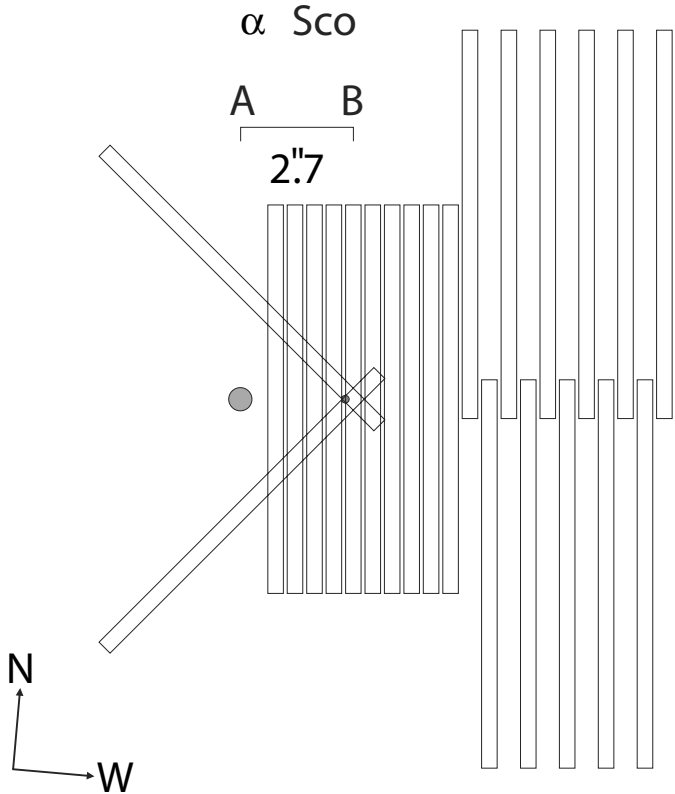}}
\caption{ Geometry of the Antares system together with true
spectrograph slit sizes and positions projected to scale on the sky.
The slit length is $10\arcsec$ referring to the blue arm.
Slit positions are $0\farcs9$, $1\farcs4$. in steps of $0\farcs5$ 
(cf Fig.~\ref{halp_emiss}) relative to the M star, perpendicular to the
line A-B. Two slit positions are tilted by 45{\degr} as displayed. 
}
\label{slitpos}
\end{figure}

The observations were performed with the UVES spectrograph
at the VLT/ESO (Dekker et al. \cite{dek00}) in service mode.
 To cover the spectral
 range from $\sim 3050$ to 11\,000~{\AA} two pairs of standard settings of
the double-spectrograph were used: one  with a central wavelength
of 3460~{\AA} for  the blue arm covering the range from $\sim 3050$ to 3850~{\AA}
and simultaneously for the red arm with a central wavelength of 5800~{\AA}
covering the range from 4200 to 6800~{\AA} with an exposure time of 100~s
and the other with an exposure time of 50~s at a central wavelength
 of 3900~{\AA} for the range from 3750 to 5100~{\AA} in the blue and
 at 8600~{\AA}
 for the range from 6600 to 11\,000~{\AA} in the red. The slit size in
 the blue arm
 was $0\farcs4 \times 10\arcsec$ and $0\farcs4 \times 12\arcsec$ in the red.
 We chose 23 different pointings for both setting pairs
 to cover the main part of the nebula in the sky (cf. Fig.~\ref{slitpos}).
 The 45{\degr} positions were observed on September
 29, 2005, the inner 10 ones on February 17, 2006, and the remaining most
 westward positions on February 16, 2006. The typical seeing was
 $0\farcs6$,
slightly larger than the slit separations.

\subsection{Data reduction}
The position of the echelle orders were determined by searching for the exposed
areas of the flat field exposures and corresponding masks were used for
 further data reduction. The flat-field corrected science
data were aligned parallel to the order boundaries and calibrated
with Th-Ar comparison exposures resulting in
two-dimensional rectangular data arrays for each order
 with one axis representing
 the wavelength, the other the position along the slit.

Unexpectedly, all spectra -- even $10\arcsec$ from the M
supergiant -- are completely corrupted by scattered light from the M star.
This was a surprise also for the UVES team at ESO. ESO finally performed a
test with the UVES slit $10\arcsec$ apart east and west
 of a bright star - with
the same result as in our spectra. Apparently, light from the bright star
is scattered by the slit jaws back to the slit viewing camera and from there
again into the spectrograph slit. The UVES team comment:
 ``Our conclusion is that
 one should not trust any light excess in the
reconstructed image, since these are most likely due to reflexions
caused by the bright star, as it is the case in the test.''

Due to the ubiquitous scattered light, no general background
reduction was possible. For each spectral line a local background had to
be determined. Due to the lack of a pure spectrum of $\alpha$~Sco~A the central
 region of the data at the first slit position -- where the scattered
light of the M star is brightest
and therefore the ratio of the nebula emission to the M star light is
smallest -- was used as a background template.
The template was fitted to the data allowing a scale and an offset. A region
of +/- 30\,km\,s$^{-1}$ around the laboratory wavelength of the spectral line was
excluded while fitting. This background was subtracted from the data. For
slit positions near the B-star additional scattered light was present. This
could be reduced using a square spline fit to the remaining continuum.

 As a further consequence of the heavy contamination of the nebula spectrum
with M star light -- the nebula spectrum is always faint compared to the scattered
light spectrum -- we are not able to provide quantitative line fluxes, except for
relative fluxes of neighboring lines.

\begin{table}[t]
\caption[]{\label{emisslines}
Antares nebula emission lines
}
\begin{center}
\begin{tabular}{lllll}
$\lambda$~~[\AA] & ion &~~~~& $\lambda$~~[\AA] & ion \\
\noalign{\smallskip}\hline\noalign{\smallskip}
3076.07 & [\ion{Ni}{ii}]     & &  4276.83 & [\ion{Fe}{ii}]   \\
3175.38 & [\ion{Fe}{ii}]     & &  4287.40 & [\ion{Fe}{ii}]   \\
3177.53 & \ion{Fe}{ii}       & &  4305.90 & [\ion{Fe}{ii}]   \\
3186.74 & \ion{Fe}{ii}       & &  4319.62 & [\ion{Fe}{ii}]   \\
3192.91 & \ion{Fe}{ii}       & &  4326.24 & [\ion{Ni}{ii}]   \\
3193.43 &                    & &  4340.47 & H$\gamma$        \\
3193.80 & \ion{Fe}{ii}       & &  4346.85 & [\ion{Fe}{ii}]   \\
3196.07 & \ion{Fe}{ii}       & &  4352.78 & [\ion{Fe}{ii}]   \\
3210.04 & \ion{Fe}{ii}       & &  4358.37 & [\ion{Fe}{ii}]   \\
3213.31 & \ion{Fe}{ii}       & &  4359.34 & [\ion{Fe}{ii}]   \\
3214.67 & \ion{Fe}{ii}       & &  4372.43 & [\ion{Fe}{ii}]   \\
3223.15 & [\ion{Ni}{ii}]     & &  4382.75 & [\ion{Fe}{ii}]   \\
3227.74 & \ion{Fe}{ii}       & &  4413.78 & [\ion{Fe}{ii}]   \\
3255.89 & \ion{Fe}{ii}       & &  4416.27 & [\ion{Fe}{ii}]   \\
3277.55 & [\ion{Fe}{ii}]     & &  4452.11 & [\ion{Fe}{ii}]   \\
3376.20 & [\ion{Fe}{ii}]     & &  4457.95 & [\ion{Fe}{ii}]   \\
3378.16 & [\ion{Ni}{ii}]     & &  4474.91 & [\ion{Fe}{ii}]   \\
3387.09 & [\ion{Fe}{ii}]     & &  4488.75 & [\ion{Fe}{ii}]   \\
3438.89 & [\ion{Ni}{ii}]     & &  4492.64 & [\ion{Fe}{ii}]   \\
3441.00 & [\ion{Fe}{ii}]     & &  4514.90 & [\ion{Fe}{ii}]   \\
3442.04 &                    & &  4728.07 & [\ion{Fe}{ii}]   \\
3452.31 & [\ion{Fe}{ii}]     & &  4774.74 & [\ion{Fe}{ii}]   \\
3455.11 & [\ion{Fe}{ii}]     & &  4814.55 & [\ion{Fe}{ii}]   \\
3501.63 & [\ion{Fe}{ii}]     & &  4861.33 & H$\beta$         \\
3504.02 & [\ion{Fe}{ii}]     & &  4874.49 & [\ion{Fe}{ii}]   \\
3504.51 & [\ion{Fe}{ii}]     & &  4889.63 & [\ion{Fe}{ii}]   \\
3514.05 &                    & &  4905.35 & [\ion{Fe}{ii}]   \\
3538.69 & [\ion{Fe}{ii}]     & &  4923.92 & \ion{Fe}{ii}     \\
3539.19 & [\ion{Fe}{ii}]     & &  5018.44 & \ion{Fe}{ii}     \\
3559.41 & [\ion{Ni}{ii}]     & &  5041.03 & \ion{Si}{ii}     \\
3626.89 & [\ion{Ni}{ii}]     & &  5055.98 & \ion{Si}{ii}     \\
3856.02 & \ion{Si}{ii}       & &  5056.32 & \ion{Si}{ii}     \\
3862.59 & \ion{Si}{ii}       & &  5111.63 & [\ion{Fe}{ii}]   \\
3970.07 & H$\epsilon$        & &  5158.78 & [\ion{Fe}{ii}]   \\
3993.06 & [\ion{Ni}{ii}]     & &  5163.95 & \ion{Fe}{ii}     \\
4101.74 & H$\delta$          & &  5169.03 & \ion{Fe}{ii}     \\
4114.48 & [\ion{Fe}{ii}]     & &  5261.62 & [\ion{Fe}{ii}]   \\
4147.26 & \ion{Fe}{ii}       & &  5273.35 & [\ion{Fe}{ii}]   \\
4177.21 & [\ion{Fe}{ii}]     & &  5333.65 & [\ion{Fe}{ii}]   \\
4178.95 & [\ion{Fe}{ii}]     & &  6347.11 & \ion{Si}{ii}     \\
4211.10 & [\ion{Fe}{ii}]     & &  6371.37 & \ion{Si}{ii}     \\
4243.98 & [\ion{Fe}{ii}]     & &  6548.05 & [\ion{N}{ii}]    \\
4244.81 & [\ion{Fe}{ii}]     & &  6562.85 & H$\alpha$        \\
4248.83 &                    & &  6583.45 & [\ion{N}{ii}]    \\
4251.44 & [\ion{Fe}{ii}]?    & &  7377.83 & [\ion{Ni}{ii}]   \\
\noalign{\smallskip}\hline
\end{tabular}
\end{center}
\end{table}

\begin{figure*}
  \resizebox{\hsize}{!}{\includegraphics*{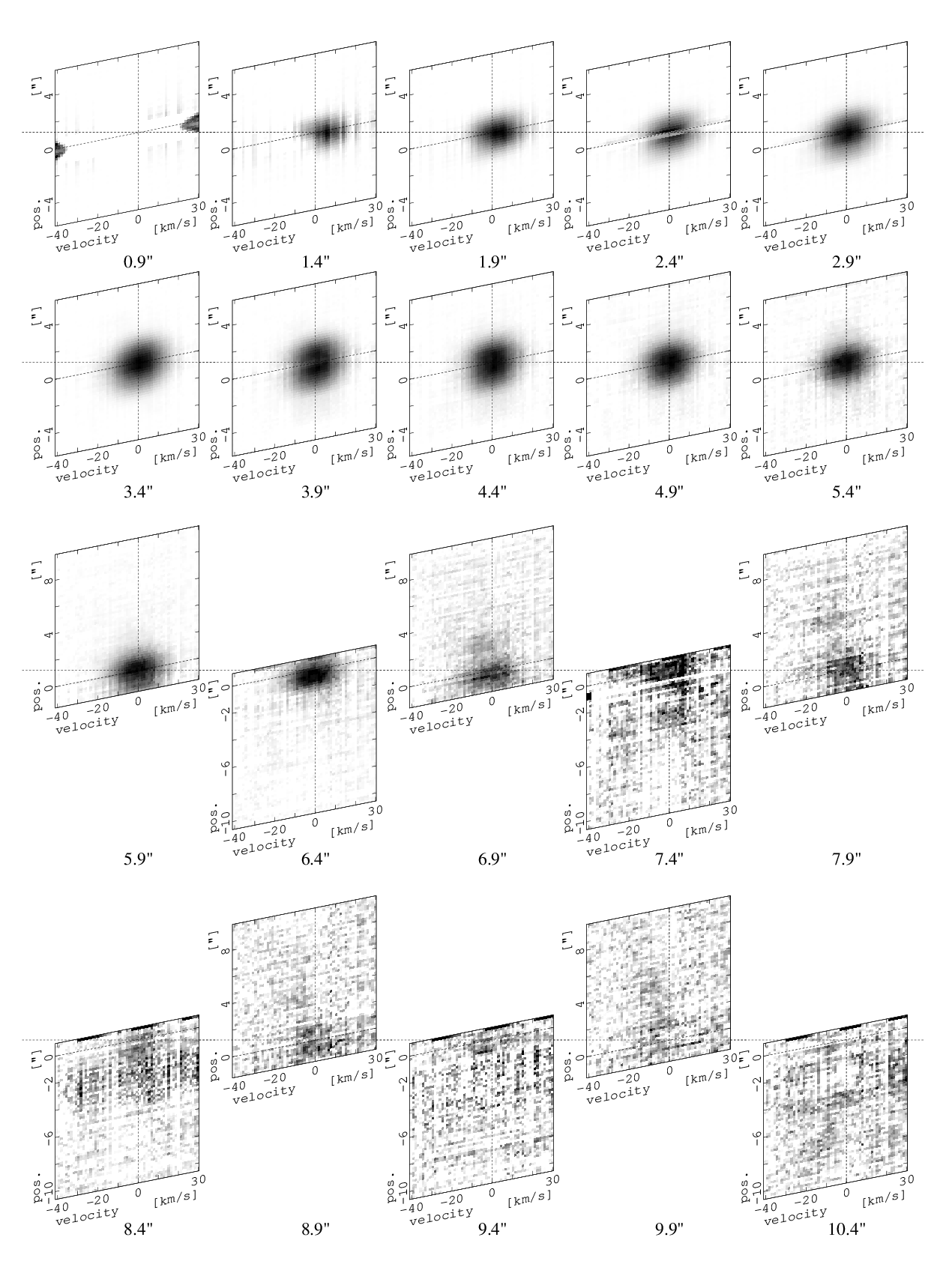}}
\caption{
H$_{\alpha}$ 2-D brightness distributions as a function of the
spectrograph slit position relative to $\alpha$ Sco A as shown in 
Fig~\ref{slitpos}.
 Notice that due to the M star wind
expansion, the velocity coordinate corresponds to the spatial depth
(perpendicular to the plane of the sky). The density in the plots has
 been rescaled in order to provide maximum visibility. The total counts
 vary by a factor of $\sim 150$ between $1\farcs9$ and $8\farcs9$.
}
\label{halp_emiss}
\end{figure*}

\begin{figure}
  \resizebox{\hsize}{!}{\includegraphics*{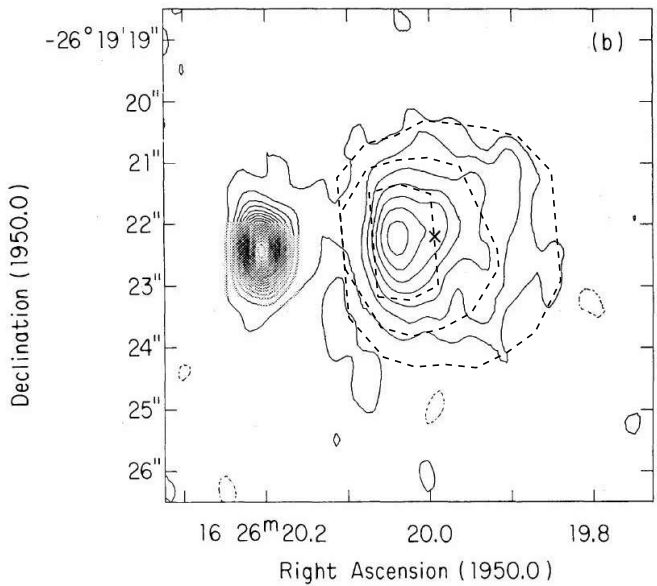}}
\caption{ Comparison of the distribution of H$_{\alpha}$ emission 
(dashed) with radio \mbox{free-free} emission (full lines)
(Hjellming \& Newell \cite{hje83}).
Dashed lines represent half, fourth, and tenth contour levels of the
 central emission flux. }
\label{hii_region}
\end{figure}

\section{Phenomenological description of the observations}
\subsection{The emission line spectrum}
A list of emission lines in the spectral range 3050 to 7400~{\AA}
is given in Table \ref{emisslines}. Compared to the line list of Swings \&
Preston (\cite{swi78}) we cover a larger wavelength range with
many new lines below 3225~{\AA} and longward of 4500~{\AA}. In
addition, the higher spectral resolution combined with a digital
detector (compared to Swings \& Preston's photographic spectra) allowed
a better subtraction of the contaminating M star light which led to
the removal of some of the lines given by Swings \& Preston,
e.g. 3427, 3991, 4018, 4088, 4127, 4178, 4233, 4423, 4471, and 4483~{\AA}.
Most of them had been classified already as doubtful by these authors.
Altogether the number of unambiguously detected emission
lines has been more than doubled. Identifications were made using
the NBS online data bank NIST (Ralchenko et al. \cite{ral07}).

For the physical interpretation of the nebula, the detection of
H$_{\alpha}$ to H$_{10}$ is important, since the recombination lines
can be used to map the geometrical extent of the \ion{H}{ii} region for
a comparison with a radio map and to locate the \ion{H}{ii} region in
velocity space which should allow us to determine its position
relative to the plane of the sky.

We confirm that the strongest lines are the forbidden iron lines
[\ion{Fe}{ii}] 4287, 4359, 4416, 4414, and 4814~{\AA}. We have also detected a
number of allowed \ion{Fe}{ii} lines. The \ion{Fe}{ii} mult.\ 42 lines
5169, 5018, and 4924~{\AA} which feed the upper level
of the strongest forbidden [\ion{Fe}{ii}] lines 4287 and 4359~{\AA} 
are noticable.
The mult.\ 42 lines are obviously produced by UV pumping
via the strong resonance scattering line of the UV mult. 3 at 2343~{\AA}
seen in UV spectra of $\alpha$ Sco B (Hagen et al. \cite{hag87}).

\begin{figure*}
  \resizebox{\hsize}{!}{\includegraphics*{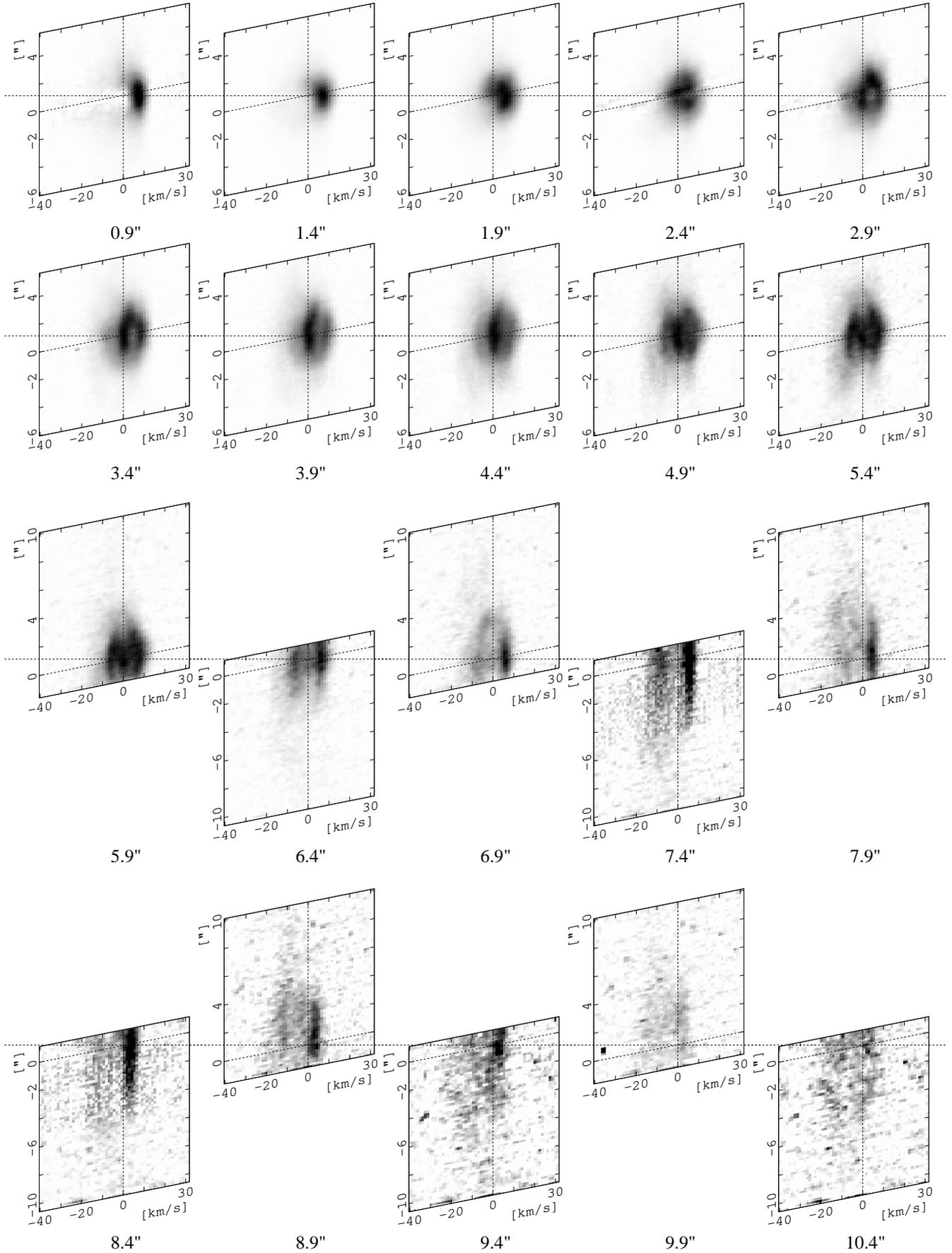}}
\caption{
[\ion{Fe}{ii}] 4814.55~{\AA} 2-D brightness distributions as a function of
the spectrograph slit position (see notes in caption of Fig.~\ref{halp_emiss}).
}
\label{fe_emiss}
\end{figure*}

\subsection{The spatial extent and velocity field in the Antares nebula}
After having carefully subtracted the underlying continua of scattered
light, the mapping of the nebula by long-slit spectroscopy in
steps of 0$\farcs$5 in the E-W direction allows us to construct 2-D images
(space versus velocity) of the emission nebula, in particular in the
strongest lines as a function of the slit position. In the following we discuss
the velocity-space images separately for
different types of lines.

\subsubsection{Balmer lines}
Previous work (Kudritzki \& Reimers \cite{kud78}; Hjellming \&
Newell \cite{hje83}) has shown that the hot B star creates an \ion{H}{ii}
region within the wind of the M supergiant. The extent of the \ion{H}{ii}
region (ionization bounded or not) depends on the mass-loss rate of the
supergiant and on the number of Lyman continuum photons of $\alpha$~Sco~B.
The \ion{H}{ii}
region was seen in free-free radio emission and its recombination
spectrum should show Balmer lines in emission.

Swings \& Preston (\cite{swi78}) did not see H$_{\beta}$ and
H$_{\gamma}$ emission lines in their photographic spectra.
Fig.~\ref{halp_emiss}
shows that H$_{\alpha}$ emission has an extent along the slit of
$\sim 4\arcsec$, nearly independent of the slit position, except the
position closest to the M star ($1\farcs4$ from the M giant) where the
emission region appears narrower ($\sim 3\arcsec$).  At position
$0\farcs9$ from the M star there is no H$_{\alpha}$ emission. In velocity space
the center of the H$_{\alpha}$ emission is always positive relative to
 the M star velocity,  indicating that the
B star is located beyond the plane of the M star (see below) in agreement with
Baade \& Reimers (\cite{baa07}).

A comparison of the extent of radio free-free emission observed with
the VLA (Hjellming \& Newell \cite{hje83}) with the extent of
H$_{\alpha}$ emission measured from the data displayed in
Fig.~\ref{hii_region} shows that around the B star and in the
direction of the M supergiant the location of H$_{\alpha}$,
H$_{\beta}$, and radio free-free emission is identical within the
resolution. West of B, H$_{\alpha}$ is more extended, since it can be
seen up to $4\arcsec$ in this direction. Apparently, the Balmer lines
are a more sensitive probe of the \ion{H}{ii} region. The Balmer line
series can be seen up to H$_{10}$.

\begin{table}[h!tb]
\caption[]{\label{lineratio} Comparison of line strengths of
H$_{\alpha}$ and [\ion{N}{ii}] 6583.45~{\AA} for slit positions in the
\ion{H}{ii} region.  }
\begin{center}
\begin{tabular}{cc}
distance to M~star~[\arcsec] & H$_{\alpha}$/[\ion{N}{ii}] \\
\noalign{\smallskip}\hline\noalign{\smallskip}
1.4 & $2.7 \pm 0.9$ \\
1.9 & $3.9 \pm 1.7$\\
2.4 & $3.7 \pm 1.6$\\
2.9 & $3.3 \pm 1.0$\\
3.4 & $3.8 \pm 1.0$\\
3.9 & $4.5 \pm 1.3$\\
4.4 & $5.8 \pm 1.9$\\
4.9 & $8.0 \pm 1.8$\\
5.4 & $10.3 \pm 5.4$\\
5.9 & $7.7 \pm 3.3$\\
6.4 & $11.5 \pm 6.5$\\
6.9 & $9.5 \pm 10.7$\\
\noalign{\smallskip}\hline\noalign{\smallskip}
\end{tabular}
\end{center}
\end{table}

\subsubsection{[\ion{N}{ii}] emission}
In all previous investigations of the Antares nebula the question
remained open, why the ``classical'' emission lines, normally
prominent in \ion{H}{ii} region spectra, like [\ion{O}{iii}] 4959/5007~{\AA}, [\ion{O}{ii}] 3726/3729~{\AA},
[\ion{N}{ii}] 6583/6548~{\AA}, and [\ion{S}{ii}] 6716/6731~{\AA}, were not detected. Our UVES spectra
allow a more critical assessment than earlier photographic spectra
like those used e.g. by Swings \& Preston (\cite{swi78}).  In
addition we have 10 long-slit positions covering the \ion{H}{ii} region
(cf. Fig.~\ref{slitpos}) which gave us the possibility of detecting faint \ion{H}{ii}
emission lines. The result is that except for [\ion{N}{ii}] 6583~{\AA} and 6548~{\AA}
none of the \ion{H}{ii} region lines, in particular [\ion{O}{ii}] 3726/3729~{\AA}
 or [\ion{O}{i}] 6300/6363~{\AA},
were detected. [\ion{N}{ii}] 6583~{\AA} closely follows the H$_{\alpha}$
distribution shown in Fig.~\ref{halp_emiss}. Since it lies close to the
H$_{\alpha}$ line, we
can measure the H$_{\alpha}$/[\ion{N}{ii}] ratio as a function of the location in
the nebula (Table \ref{lineratio}).

\begin{figure}[h]
  \resizebox{\hsize}{!}{\includegraphics*{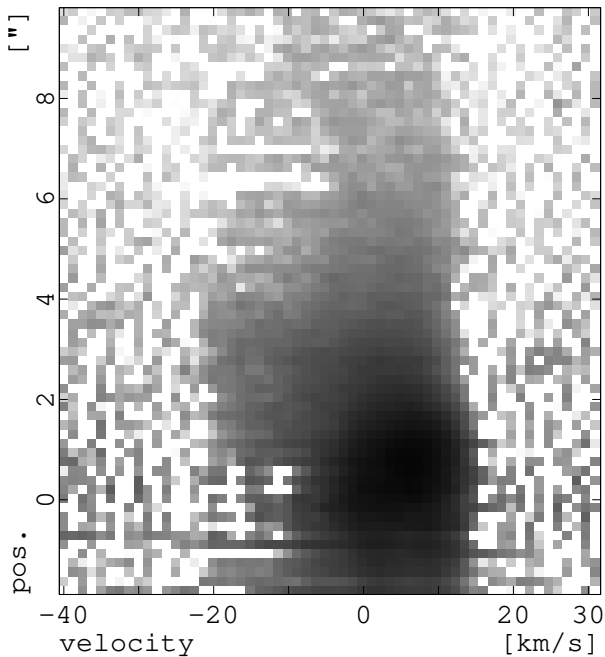}}
\caption{
[\ion{Fe}{ii}] 4814.55~{\AA} 2-D brightness distribution at the 45{\degr}
 NE slit position, logarithmically scaled.
}
\label{fe45deg}
\end{figure}

\subsubsection{The allowed lines}
The \ion{Si}{ii} lines 6347, 6371, 3862, and 3856~{\AA} roughly follow 
the H$_{\alpha}$ distribution (Fig.~\ref{halp_emiss}) in that the emission
does not
extend beyond 1$\farcs$5 east of B. In velocity space they are
narrower (0 to +10 km\,s$^{-1}$) than the H$_{\alpha}$ distribution.
 The reason cannot be the smaller thermal broadening of \ion{Si}{ii}
lines compared to H$_{\alpha}$, since at 5000~K the thermal width is
$\sim 1\,{\rm km\,s^{-1}}$, small compared to the much larger wind
microturbulence that is typically half the wind velocity. The
simplest interpretation would be that the \ion{Si}{ii} lines are formed close
to the B star by \ion{Si}{iii} recombination in a narrower velocity space
(wind sector) than H$_{\alpha}$. Unfortunately we have no \ion{Si}{iii} 1206~{\AA}
absorption data. However, the existence and velocity of \ion{Al}{iii}
absorption lines do show that they must be formed in the immediate
surroundings of the B star (Baade \& Reimers \cite{baa07}).  The
narrowness of the \ion{Si}{iii} region is confirmed by our final \ion{H}{ii}
region model.

The strongest allowed line is \ion{Fe}{ii} 3228~{\AA}. Its
distribution is similar to that of the [\ion{Fe}{ii}] lines and distinctly
different from \ion{Si}{ii}. It is clearly present 0$\farcs$9 from the M
supergiant, i.e. outside the \ion{H}{ii} region. The most interesting allowed
\ion{Fe}{ii} line is 5169~{\AA}, since the appearance of this line
both in the neutral wind (position 0$\farcs$9 from the M star) and in the \ion{H}{ii}
region indicates that one channel for the excitation of the forbidden
[\ion{Fe}{ii}] lines is line scattering of B star photons in the strong UV
\ion{Fe}{ii} resonance lines. One of the strong UV resonance lines in the UV
spectrum of $\alpha$ Sco B is the UV mult.\ 3 line at 2343~{\AA}. These lines show
P Cyg type profiles (cf.\ Hagen et al. \cite{hag87}).  In case of
the 2343~{\AA} line a second downward channel exists via the
\ion{Fe}{ii} lines 4924, 5018, and 5169~{\AA}. The branching ratio of the 2343~{\AA}
and 4924 - 5169~{\AA} lines is $\sim 16$ (NIST-data base,
Ralchenko et al. \cite{ral07}), so that
part of the downward cascade is via the three 4923 - 5169~{\AA} lines and
this is what we observe. In the neutral wind we can detect both the
2343~{\AA} reemission part of the P Cyg profile and the
4924 - 5169~{\AA} channel in emission which populates the
upper level of the strongest [\ion{Fe}{ii}] lines 4287~{\AA} and 4359~{\AA}.

\begin{figure*}
  \resizebox{\hsize}{!}{\includegraphics*{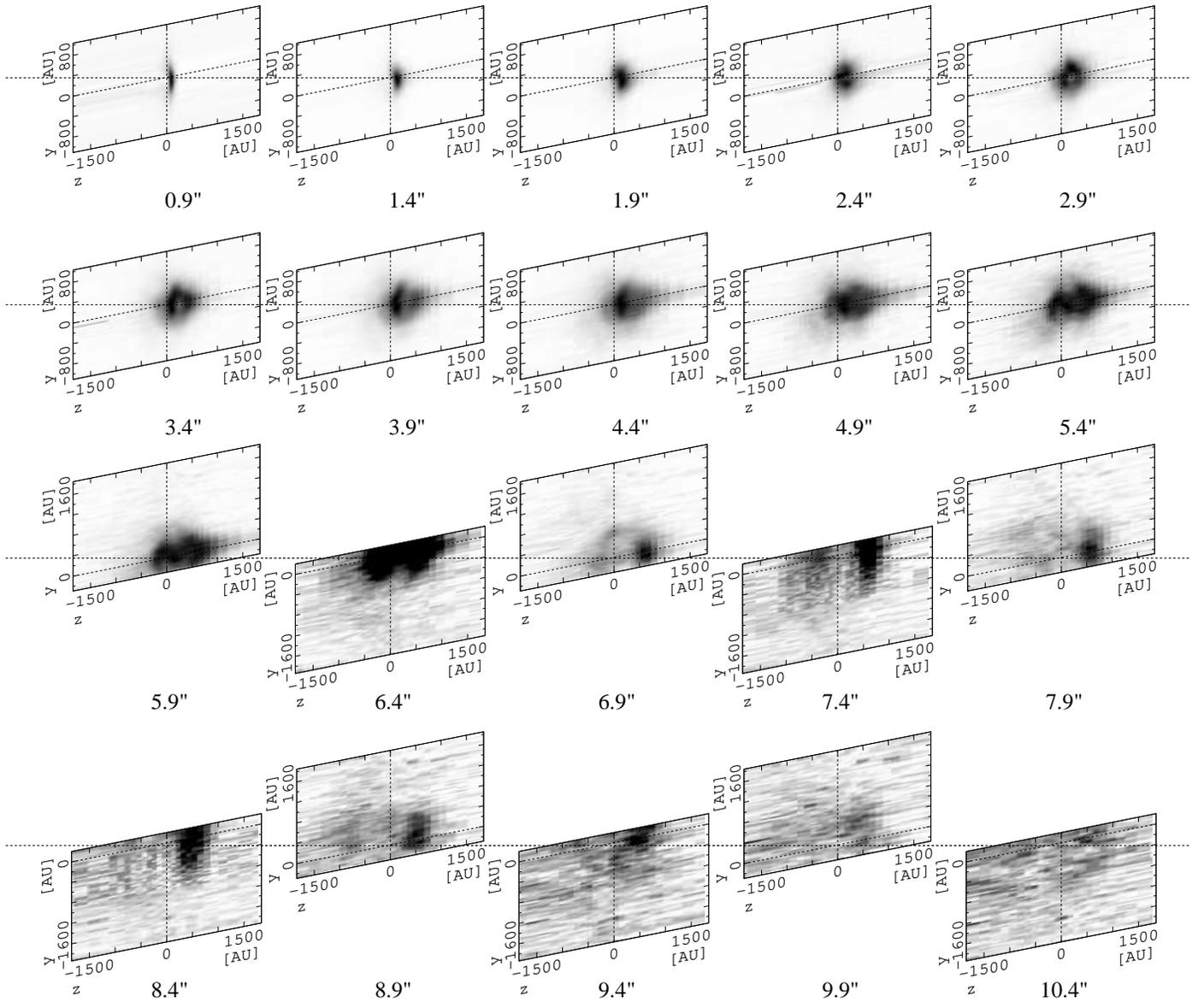}}
\caption{ [\ion{Fe}{ii}] 4814.55~{\AA} 2-D brightness distributions of
Fig. \ref{fe_emiss} transformed to spatial coordinates using an M~star
wind velocity of 20~km\,s$^{-1}$.  }
\label{fe_emiss_yz}
\end{figure*}

In conclusion: We have directly identified  the [\ion{Fe}{ii}] excitation
mechanism in the cold neutral M supergiant wind, i.e. the faint
extended emission region outside the \ion{H}{ii} region which is marked by
H$_{\alpha}$ emission, radio free-free emission, and \ion{Si}{ii}
emission. This \ion{Fe}{ii} resonance line scattering mechanism is certainly
also operating in the \ion{H}{ii} region, but the comparison of the
distribution of the strong [\ion{Fe}{ii}] lines with \ion{Fe}{ii} 5169~{\AA} shows
that within the \ion{H}{ii} region either other excitation mechanisms are
dominant, e.g. collisional excitation by free electrons in the \ion{H}{ii}
region, or iron is more abundant in the gas phase.

\subsubsection{[\ion{Fe}{ii}] emission}
In the most extensive study of the emission spectrum of the Antares
nebula by Swings \& Preston (\cite{swi78}), the ``[\ion{Fe}{ii}] rich''
nebula was described as an asymmetrical region of about 3$\farcs$5
around the B star with strong lines, surrounded by a much more
extended region of weaker lines. Our data, shown in Fig.~\ref{fe_emiss},
 are largely consistent with Swings \& Preston
(\cite{swi78}) but give a more detailed picture due to the higher
spectral resolution (80\,000) and angular resolution, since most of the
data were taken at 0$\farcs$6 seeing. This means that the
resolution corresponds roughly to the slit steps (Fig.~\ref{slitpos}). Below
we summarize the main characteristics:

\begin{itemize}

\item We notice in
particular that [\ion{Fe}{ii}] emission is apparently concentrated on the
ionization front in a cone-like structure. This can be seen clearly in
Fig.~\ref{fe_emiss} at slit positions 2$\farcs$9, 3$\farcs$4, 5$\farcs$9,
and 6$\farcs$9. The reason for the [\ion{Fe}{ii}] enhancement at the \ion{H}{ii}/\ion{H}{i}
interface may be shock heating. Since the pressure in the \ion{H}{ii} region
is higher by a factor of $\sim 10$ than in the cool wind, a shock
front will build up at the interface (cf.\ Kudritzki \& Reimers
\cite{kud78}). The occurrence of [\ion{Fe}{ii}] in the Orion Nebula has
also been observed close to the ionization front (Bautista et
al. \cite{bau94}).

\item Contrary to the nondetection in
H$_{\alpha}$, [\ion{Fe}{ii}] is already seen 0$\farcs$9 from the M
supergiant. This is a consequence of the origin of these lines that are
produced via pumping processes as described in section 3.2.3 and 4.5.

\item It is obvious that beyond $\sim 5\farcs$9 west of A, the
[\ion{Fe}{ii}] emission, although weak, becomes more extended and structured
(``cloudy'').

West of B and increasing with distance from B the
lines appear as double and triple. The double lines were already reported
 by Deutsch (\cite{deu60}) and Struve \& Zebergs (\cite{str62}).

Both of the 45{\degr} positions with slit length 12$\arcsec$
 starting from B (Fig.~\ref{fe45deg}) show that the emission
 maximum is not close to B but
$\sim 1\arcsec$ apart. Furthermore, weak [\ion{Fe}{ii}] emission is seen all
along the slit, confirming Swings \& Preston (\cite{swi78}) who
saw [\ion{Fe}{ii}] up to 15$\arcsec$ in the SE direction.

 \item It can also be seen from Fig.~\ref{fe_emiss} that in the emission
 line region the median in velocities  moves systematically from
 $\sim 4$ km\,s$^{-1}$ 1$\farcs$9 west of A over 3 km\,s$^{-1}$ at B
 to 0 km\,s$^{-1}$ 5$\farcs$9 west of A and -5 km\,s$^{-1}$ at
 9$\farcs$9 west of A.

\item The nebula shows large scale deviations from a simple
geometry like a smooth spherically expanding M star wind in which
the B star produces an ellipsoid-like \ion{H}{ii} region. In reality the [\ion{Fe}{ii}]
emitting regions have a clumpy structure with discrete ``blobs''. This
is in accordance with the shell structure seen in the IR (Marsh et
al. \cite{mar01}) and the episodic mass loss seen in UV resonance
lines (Baade \& Reimers \cite{baa07}).
\end{itemize}

\section{Analysis and discussion}

\subsection{The location of B relative to the plane of the sky}
The velocity of the circumstellar \ion{Ti}{ii} absorption lines in the
line of sight had been used together with the assumption of a
spherical wind with a constant expansion velocity to estimate a
position of B $\sim 600$ AU in front of the plane of the sky
(Kudritzki \& Reimers \cite{kud78}). However, the observed episodic
nature of mass loss of $\alpha$ Sco A (4 absorption components)
disagrees with these assumptions and the derived location must be
considered as erroneous (Baade \& Reimers \cite{baa07}).

GHRS/HST spectra show that the \ion{Al}{iii} absorption lines at +7.8 km\,s$^{-1}$,
which must be formed close to the B star, favor a location of B
roughly 224 AU behind the M star (Baade \& Reimers \cite{baa07}). We
can determine the B star position independently using the velocity of
the H$_{\alpha}$ line at the eastern boundary of the \ion{H}{ii}
region at the slit positions 1$\farcs$4 and 1$\farcs$9.
The basic underlying assumption is that the M star wind velocity
is not severely disturbed and that the width of the emission defines the
front and rear side of the \ion{H}{ii} region.
The mean observed velocity in the
line of sight is $\sim 4.8$ km\,s$^{-1}$, i.e.  7.8 km\,s$^{-1}$ relative to the M
supergiant. Using a constant M star  wind velocity of 20 km\,s$^{-1}$ (Baade \& Reimers
 \cite{baa07}), we obtain a position angle of 23$^{\circ}$
behind the M star with an uncertainty of about 5$^{\circ}$.
A measurement of the [\ion{Fe}{ii}] lines of the \ion{H}{i} region in front of the \ion{H}{ii} region,
where the lines are formed by UV pumping, gives a nearly identical result.

\subsection{Geometry and binary orbit}
Long-term observations of the visual binary have shown that
the orbit is nearly in the line of sight (Hopmann \cite{hop57}),
and the binary separation has become smaller from 3$\farcs$01 in 1930
(photographic plates, Hopmann \cite{hop57}) to 2$\farcs$86 in 1989
(interferometric measurements, McAlister et al. \cite{mca90}), and
to 2$\farcs$74 in 2005 (lunar occultation, S\^oma \cite{som06}).
The extrapolation yields a separation of 2$\farcs$73 in February 2006.
With the Hipparcos distance of 185~pc and assuming a circular orbit we obtain a semi-major axis
of 549~AU. With 18 M$_{\odot}$ and 7.2 M$_{\odot}$ for $\alpha$ Sco A and B, respectively
(Kudritzki \& Reimers \cite{kud78}), Kepler's third law yields a
binary period of 2562~yrs. The corresponding
calculated orbital velocities are $v_{\rm M} = -1.82\,\rm km\,s^{-1}$ and $v_{\rm B} =
4.56\,\rm km\,s^{-1}$, with radial components $v_{\rm M,r} = -1.68\,\rm km\,s^{-1}$ and
$v_{\rm B,r} = 4.20\,\rm km\,s^{-1}$. This is close to the observed
values of -3 km\,s$^{-1}$ and +3 km\,s$^{-1}$ for the
M and B star, respectively (Evans \cite{eva67}; Kudritzki \&
Reimers \cite{kud78}), if a small systemic velocity of -1.25 km\,s$^{-1}$
is added. Unfortunately the observational data do not allow us to establish a
reliable orbital solution. If we trust the 19th century measurements compiled by
Hopmann (\cite{hop57}) the orbit cannot be circular. However, for the present
work the uncertainties of the binary orbit can be neglected.

\subsection{The shape of the \ion{H}{ii} region}
The apparent shape of the H$_{\alpha}$ emission nebula in the sky
according to Fig.~\ref{halp_emiss} is displayed in
Fig.~\ref{hii_region}, together with the radio map of Hjellming \&
Newell (\cite{hje83}).  Details of the H$_{\alpha}$ and [\ion{Fe}{ii}]
intensity projected onto the plane of the sky (2-D) have been
presented in the sections 3.2.1 and 3.2.4. The emission nebula can be
best described by a bow-shock shaped region with its front $\sim 1
\arcsec$ east of B which opens slowly to an extent of roughly
6$\arcsec$ up to $\sim 3 \arcsec$ west of B. Beyond 3$\arcsec$ the
``shock cone'' opens with filamentary structures to an extent of up to
nearly 16$\arcsec$ ($\pm$ 8$\arcsec$). In velocity space the ``double
line'' character becomes more distinct, and typically the red line
edge is sharply limited. We appear to look through a shock cone with a
front and rear edge (cf. Fig.~\ref{fe_emiss} on and slightly west of
B).

With the assumption that the M star wind continues at constant
velocity through the \ion{H}{ii} region without being grossly disturbed, we
can construct a 3-D image of the [\ion{Fe}{ii}] emitting gas since each
velocity in Fig.~\ref{fe_emiss} corresponds in
that case to a location relative to
the plane of the sky (positive velocities behind, negative velocities
in front of the plane of the sky). The resulting 2-D slices for each
slit position are presented in Fig.~\ref{fe_emiss_yz}: East of B the
 emission comes
from behind the plane of the sky, as expected, since the B star and
the surrounding \ion{H}{ii} region are behind the plane of the sky. Between
the position of B and $\sim 3\arcsec$ west of B, the emission comes from
a region of cylindrical shape with a diameter of $\sim 6 \arcsec$. This
extent is almost identical in the plane of the sky  and
in the vertical direction. A gradual apparent
shift of the [\ion{Fe}{ii}] emitting gas is seen
 from 200~AU behind the plane at
1$\farcs$4, slightly behind the plane of the sky (on average)
 to an extent of between 740~AU behind to 1480~AU in front of the
plane of the sky. The shock cone apparently opens up beyond 3$\arcsec$ west
of B to larger distances in front of the plane of the sky.

\subsection{Where are the ``normal'' forbidden \ion{H}{ii} region lines?}
The Antares nebula has been considered as ``peculiar'' from its detection
(cf.\ the extensive discussion in Struve \& Zebergs \cite{str62})
because its spectrum is dominated by numerous strong [\ion{Fe}{ii}] lines
while the classical \ion{H}{ii} region lines [\ion{O}{ii}] 3287~{\AA}, [\ion{O}{ii}] 3726/3729~{\AA},
[\ion{O}{iii}] 4957/5007~{\AA}, and
[\ion{S}{ii}] are missing. Only the [\ion{N}{ii}] lines are present. Since $\alpha$ Sco
B is with B 2.5 V of much later spectral class than the O stars
normally exciting an \ion{H}{ii} region, we expect a lower electron
temperature. Swings \& Preston (\cite{swi78}) had estimated
$T_{\rm e}\approx 4000$\,K from the [\ion{Fe}{ii}] excitation, while Kudritzki \&
Reimers (\cite{kud78}) estimated $T_{\rm e} = 4000$\,K from the
heating/cooling balance. The latter value , however, was based on normal \ion{H}{ii}
region cooling rates dominated by [\ion{O}{iii}], [\ion{O}{ii}] etc. and is not
applicable according to the present work.

\begin{figure}
  \resizebox{\hsize}{!}{\includegraphics*{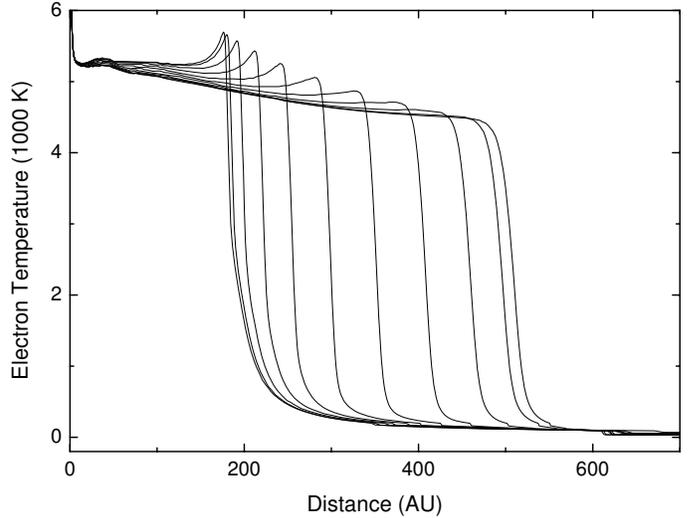}}
  \caption{
  Theoretical run of the electron temperature in different
  directions as a function of distance from the B star. The CLOUDY
  models are presented for 11 equidistant angles in the \ion{H}{ii}
  region between 0{\degr} and 180{\degr}, where 0{\degr} corresponds
  to the connecting line between the two stars. The density distribution
 used here and in Fig.~\ref{slit_int} is $n \sim r^{-2}$ according to constant
 wind expansion from the M star.
}
  \label{temperature}
\end{figure}

\begin{figure}
  \resizebox{\hsize}{!}{\includegraphics*{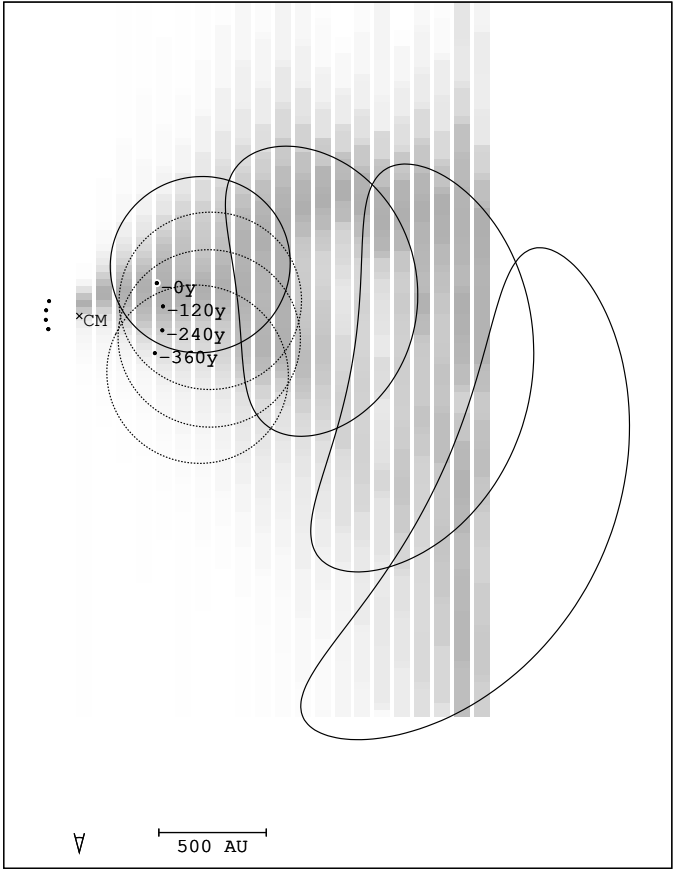}}
\caption{ ``Look back'' contours (solid lines) of \ion{H}{ii} region
borders in a plane perpendicular to the plane of the sky caused by the
combination of wind expansion and orbital motion of $\alpha$ Sco B
relative to $\alpha$ Sco A shown for present and previous phases (-120
yrs to -360 yrs; 130 years being the wind travel time for the
projected distance between A and B). This can be interpreted as a qualitative model for the
comet like structure of the [\ion{Fe}{ii}] emission regions. For clarification
 we overlay
the projected brightness distribution of Fig.~\ref{fe_emiss_yz}.
 }
\label{hii_hist}
\end{figure}

The key to an understanding of the missing [\ion{O}{ii}],
[\ion{O}{iii}] lines with the simultaneous presence of the
[\ion{N}{ii}] lines is twofold: electron temperature and abundance. At
first, the electron temperature of the \ion{H}{ii} region must be so
low that for normal or slightly reduced O/H ratios [\ion{O}{ii}]
3726/3729~{\AA} must be below the detection limit.  For [\ion{O}{ii}]
3726/3729~{\AA} we can give an upper limit: we have detected the
Balmer line series up to H$_{10}$, H$_{11}$ (close to [\ion{O}{ii}])
not being visible. Thus the H$_{10}$ intensity, as predicted by
recombination theory, can be regarded as an upper limit to
[\ion{O}{ii}] which yields [\ion{O}{ii}]/H$_{\alpha} <$ 0.017 for 3726~{\AA}.
Models calculated for the Antares \ion{H}{ii} region with the ionization code
CLOUDY (version 07.02.00, last described by Ferland et al.\
\cite{fer98}) lead to $T_{\rm e} < 5000$\,K. At such a low $T_{\rm e}$,
[\ion{N}{ii}]/H$_{\alpha}$ would be $< 0.1$ (see also Fig. 3 in
Mierkewicz et al. \cite{mie06}) for solar abundances and thus below
our detection limit, while the observed [\ion{N}{ii}]
6583~{\AA}/H$_{\alpha}$ is between 0.3 and 0.4 in the central region
of the \ion{H}{ii} region. The solution to this puzzle is that the
atmospheric (and wind) composition of the M supergiant must have been
altered by the CNO cycle.

When a star ascends the red giant branch the first dredge-up may alter
the CNO abundance significantly. The exact amount of the depletion of
O and enhancement of N depends on the initial mass. As a consequence
it is difficult to derive exact CNO abundances for $\alpha$ Sco A on
the basis of evolutionary tracks, since neither its current mass nor
the mass it had on the main sequence are reliably known.  Recently
Balachandran et al. (\cite{bal06}) have published abundance
measurements of a number of M type supergiants that suggest $\rm [N/O] \sim 0.7$
for $\alpha$ Sco A.

\begin{figure*}[h!tb]
\resizebox{\hsize}{!}{\includegraphics*{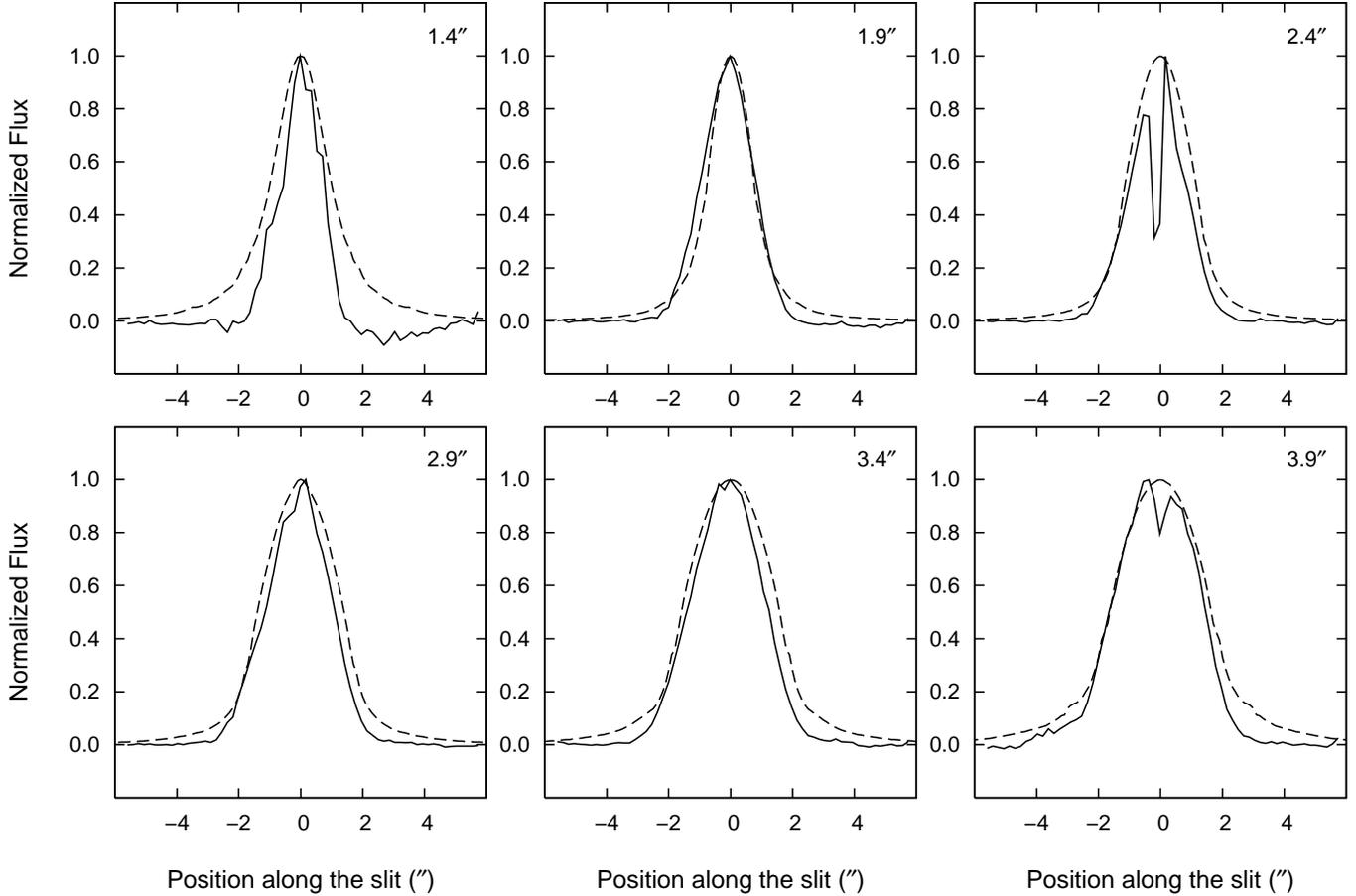}}
\caption{Comparison of the velocity-integrated H$_\alpha$ flux
 with CLOUDY
simulations shown for a mass-loss rate of $1.05 \times 10^{-6}$
M$_{\odot}$\,yr$^{-1}$.  Since we have no absolute fluxes, both the
observed (solid line) and the theoretical fluxes (dashed line) are
normalized.}
\label{slit_int}
\end{figure*}

While the CNO abundance ratios have not been measured in $\alpha$ Sco
A, Harris \& Lambert (\cite{har84}) have found that $^{17}$O and
$^{13}$C are enriched so that there is also clear observational
evidence for CNO cycle material in the envelope of $\alpha$ Sco A. Our
best model for the \ion{H}{ii} region leads to $\rm N/N_\odot = 3.3
\pm 1$ and $\rm O/O_\odot \la 0.9 \pm 0.1$, i.e $\rm [N/O] = 0.56 \pm
0.2$ which is consistent with the above-mentioned values for M
supergiants. The final CLOUDY \ion{H}{ii} region model used for 
Fig~\ref{temperature} and Fig~\ref{slit_int} has solar abundances except
for CNO for which we used the above numbers.

The nondetection of [\ion{S}{ii}] 6716/6731~{\AA} can be understood
with the same arguments. In classical, relatively cool \ion{H}{ii}
regions, [\ion{S}{ii}]/[\ion{N}{ii}] is typically $\sim 0.3$ for solar
abundances. With N being enhanced by a factor of more than 3, the
expected [\ion{S}{ii}]/[\ion{N}{ii}] ratio is $< 0.1$, consistent with
the nondetection in our spectra.

The ionization structure and temperature of the \ion{H}{ii} region
depend primarily on the number of incident Lyman continuum photons
emitted by the B star. The most reliable estimate of the Lyman
continuum luminosity $L_{\rm Ly}$ comes from radio measurements of the
\ion{H}{ii} region by Hjellming \& Newell (\cite{hje83}). It is
possible to adjust the effective temperature of the B star until the
measured radio flux is reproduced. These calculations have to be
performed iteratively, since the electron temperature $T_{\rm e}$ of
the \ion{H}{ii} region is a priori unknown. Only the knowledge of
$T_{\rm e}$ allows us to quantify the relation between the Lyman
continuum luminosity and the radio flux. A final model that satisfies
all observational constraints yields a mean electron temperature of
4900~K. A detailed plot of the temperature distribution in the
\ion{H}{ii} region is shown in Fig.~\ref{temperature}. It should be noted
that \ion{Fe}{ii} has a considerable impact on the \ion{H}{ii} region and its
electron temperature. As a consequence we apply the CLOUDY code with the more
accurate \ion{Fe}{ii} model using 371 levels as described by Verner et al.
(\cite{ver99}).

\subsection{What excites the [\ion{Fe}{ii}] lines?}
The [\ion{Fe}{ii}] lines have been observed in high-density regions
such as the Orion Nebula (Bautista et al. \cite{bau94}; Baldwin et al.
\cite{bal00}) and it has been shown by multilevel
collisional-radiative \ion{Fe}{ii} models that under the Orion
Nebula conditions (at $n_{\rm e} = 10^{4}\,\rm cm^{-3}$ and $T_{\rm
e} = 10^{4}\,\rm K$) the optical [\ion{Fe}{ii}] lines (4814,
4277~{\AA} etc.) are due to UV continuum pumping, while the infrared
[\ion{Fe}{ii}] lines are produced by collisional excitation (Baldwin
et al. \cite{bal96}, Verner et al. \cite{ver00}). The strong
[\ion{Fe}{ii}] line at 4814~{\AA} as presented in
Fig.~\ref{fe_emiss} is the result of a pumping route starting from
level a$^4$F which is excited to level x$^4$F$^{\rm o}$. The
downward transition to b$^4$F followed by a transition back to the
a$^4$F term finally produces the observed line. Details of this and
other \ion{Fe}{ii} pumping processes are discussed thoroughly by
Verner et al. \cite{ver00} (see especially their Fig.~7 showing all
relevant routes of the continuum pumping).

In the case of the $\alpha$ Sco wind we explicitly showed above
that in the neutral, cool wind (\ion{H}{i} region) the
[\ion{Fe}{ii}] lines are visible and are formed by UV pumping. So the
question remains: Why is the contrast between \ion{H}{ii} region and
\ion{H}{i} region (Figs.~\ref{halp_emiss},~\ref{fe_emiss}) so strong?
 At first, the
[\ion{Fe}{ii}] lines are strongest at the interface between the \ion{H}{ii}
and the \ion{H}{i} region. Part of the explanation for this behavior
is the fact that the \ion{Fe}{iii} zone around the B star nearly fills
the \ion{H}{ii} region (in contrast to the smaller \ion{Si}{iii} zone).
This may lead to the triple velocity structure in [\ion{Fe}{ii}] clearly
visible at position $5\farcs4$ and $5\farcs9$ (Fig.~\ref{fe_emiss}),
where the central component comes from \ion{Fe}{iii} recombination and
the `edge' components from close to the ionization front.

Secondly, as has been shown by HST
spectra, iron (and consequently \ion{Fe}{ii}) is depleted onto dust
by at least a factor of 10 in the M giant wind (Fig.~4 in Baade \&
Reimers \cite{baa07}). The combination of density enhancement and
grain destruction in the shock heated zone might be a further reason for
the apparent [\ion{Fe}{ii}] line enhancement at the
\ion{H}{ii}/\ion{H}{i} interface. A further argument against pure
collisional excitation of the [\ion{Fe}{ii}] lines within the
\ion{H}{ii} region is our observation that the density enhanced
shock zone is visible outwards to at least 7$\arcsec$ west of B, far
beyond the \ion{H}{ii} region which extends to roughly 3$\arcsec$
only.

The density enhancements made visible through [\ion{Fe}{ii}]
emission are apparently carried outwards by the M supergiant wind
far beyond the \ion{H}{ii} region and remain visible although
\ion{H}{ii} has recombined. Notice in particular the bright rear
side and the fainter front side of the shock cone. The wind travel
time $t_{\rm w}$ for 550~AU is $\sim 130$ yrs, to be compared with
the recombination time $t_{\rm rec}$ of roughly 18 yrs for a
mass-loss rate of $10^{-6}$ M$_{\odot}$\,yr$^{-1}$ at the western
border of the \ion{H}{ii} region ($\sim 3 \arcsec$ or $\sim 550$~AU
west of B). Though $t_{\rm rec}$ is short compared to the wind travel
time, advection effects may explain the existence of weak H$_\alpha$
emission outwards to $\sim 6 \arcsec$ west of B.

We have schematically sketched this behavior showing the combined
effect of B star orbital motion and wind expansion on the movement
of a ``former'' \ion{H}{ii} region (as seen in Fe) relative to the M
supergiant. The simple model (Fig.~\ref{hii_hist})
reproduces the location of the front and the rear edge of former \ion{H}{ii}
region gas in velocity space. With increasing distance from the B
star, the median velocity moves from $+$4 km\,s$^{-1}$ 1$\arcsec$ E of B over
$+$3 km\,s$^{-1}$ at B to 0 km\,s$^{-1}$ W of B and $-$5 km\,s$^{-1}$ at 7$\arcsec$ W of B,
consistent with the -- vertical to the plane of the sky -- cometary
shaped \ion{Fe}{ii} emission region as a result of the wind-carried density
enhancements of the \ion{H}{i}/\ion{H}{ii} region interface.

From the above discussion it is obvious that a quantitative explanation
of the [\ion{Fe}{ii}] emission requires substantial additional effort: In
a final step a non-stationary hydrodynamical model of the \ion{H}{ii} region
moving with the B star through the M star wind has to be computed. In a second step
this model will be the basis for a 3D-NLTE radiative transfer model of \ion{Fe}{ii}
including the B star with the UV photons which, as we have shown, is at least
partly responsible for the excitation of the [\ion{Fe}{ii}] lines. Such a model
is beyond the scope of the present paper.

\subsection{Mass-loss rate}
To quantify the mass-loss outflow we proceeded in two steps. Following
Nussbaumer \& Vogel (\cite{nus87}) we calculated in a first step 
the shape of the
\ion{H}{ii} region as a function of the mass-loss rate assuming an
electron temperature of 4000\,K (Kudritzki \& Reimers \cite{kud78})
which is considered to be constant throughout the envelope. The
resulting theoretical \ion{H}{ii} region was projected onto the
plane of the sky and compared to the observed H$_{\alpha}$ emission
along the slits. Thus we obtained a tentative mass-loss rate of $8
\times 10^{-7}$ M$_{\odot}$\,yr$^{-1}$ that was used as the initial
value for the subsequent step. In order to achieve a self-consistent
model we used the ionization code CLOUDY (Ferland et al. \cite{fer98})
 and constructed a static
\ion{H}{ii} region considering all relevant heating and cooling
processes. The resulting temperature distribution is shown in Fig.~\ref{temperature}.
Now a more elaborate estimate of the mass-loss rate is possible using
the spatially dependent emissivity.
The emergent intensity of the H$_{\alpha}$ emission has
been integrated for all slit positions as a function of the
perpendicular distance from the central line. The best match with
the observed intensity distribution yields mass-loss rates between
$8.4 \times 10^{-7}$ M$_{\odot}$\,yr$^{-1}$ and $1.26 \times
10^{-6}$ M$_{\odot}$\,yr$^{-1}$. Fig.~\ref{slit_int} shows the
result for the mean value. There is a clear tendency that the
required mass-loss rate increases for the outer slit positions. It is, however,
questionable whether this finding reflects a true change in the mass
outflow or is only an artifact of inhomogeneities or curved wind
trajectories. It should be noted that the first slit position covers only
part of the \ion{H}{ii} region, resulting in a reduced emission.

We are aware that the static results of the CLOUDY code should be interpreted with care, but
for optically thin media a correct implementation of the microphysics is more important than
kinematic effects. Future work should include dynamical and time-dependent processes. Especially advection terms
and hydrodynamical effects have to be considered carefully. Some of the observed features are clearly
beyond the capabilities of a stationary model, as discussed above.


\end{document}